\documentclass[preprint,showpacs,aps]{revtex4}
\usepackage{graphicx}
\newcommand{\be}{\begin{equation}}
\newcommand{\ee}{\end{equation}}
\newcommand{\ben}{\begin{eqnarray}}
\newcommand{\een}{\end{eqnarray}}
\newcommand{\Slash}[1]{\!\!\not\!{#1}}

\def\bs{b\!\!\!/}
\def\ps{p\!\!\!/}

\def\g{\gamma}

\def\bb{\bibitem}
\def\half{{\textstyle{1\over 2}}}

\newcommand{\incps}[5]{\includegraphics[#2,#3][#4,#5]{#1}}

\newcommand{\remark}[1]{}
\newcommand{\fgl}[1]{\hspace{0.75cm}#1\hspace{-0.75cm}}

\begin{document}

\title{Chern-Simons-like action induced radiatively in General Relativity}

\author{T. Mariz, J. R. Nascimento, E. Passos and R. F. Ribeiro}
\email{tiago, jroberto, passos, rfreire@fisica.ufpb.br}
\affiliation{Departamento de F\'\i sica, Universidade Federal da Para\'\i ba, 
Caixa Postal 5008, 58051-970 Jo\~ao Pessoa, Para\'\i ba, Brazil}

\date{\today}

\begin{abstract}
The Chern-Simons-like gravitational action is evaluated explicitly in four 
dimensional space-time by radiative corrections at one-loop level. The
calculation is performed in fermionic sector where the 
Dirac fermions interact with 
the background gravitational field, including the parity-violating term 
$\bar\psi \bs\gamma_5\psi$.
The investigation takes into account the weak field approximation and dimensional 
regularization scheme.
\end{abstract}

\maketitle

\section{Introduction}
Theory of the electromagnetism was crucial to question Galilei invariance,
to give rise to Lorentz symmetry. Nowadays, in string theory one may find a way to 
question Lorentz invariance, since there are interactions that support spontaneous breaking 
of Lorentz symmetry\cite{kost}, one of those interactions being described by the 
Chern-Simons-like action.

The Chern-Simons term was first introduced in three-dimensional gauge field 
and gravitational field theories by Deser, Jackiw and Templeton \cite{DJT}. 
In the gauge theories interesting phenomena are exhibited
such as exotic statistics, fractional spin and massive gauge field. These
phenomena are of topological nature and they can be produced when 
we add Chern-Simons term to Lagrangian which describe the system under consideration.
Posteriorly, was
observed that if one adds the Chern-Simons-like term in four dimensional space-time
to Maxwell's theory both Lorentz and CPT symmetries are violated. The model 
predicts the rotation of the plane of polarization of radiation from distance galaxies,
an effect which is not observed yet \cite{cfj}. In recent
papers\cite{JP, L}, it has been studied the modification of general relativity 
when one adds the Chern-Simons-like gravitational term. 
The authors have observed that in this modified theory the Schwarzschild metric is 
solution, gravitational waves possess two polarizations which travel with 
the velocity of light and polarized waves are suppressed or enhanced. 

The induction by radiative correction of the Chern-Simons term has been analyzed
in the last twenty years. Redlich \cite{R} in a seminal paper studied the subject
in the context of the quantum electrodynamics in three-dimensional of space-time.
Following this paper other models in quantum field theory were investigated\cite{mva, mg, ro}.
 Extension to higher odd-dimension was done and also in the case of a 
gravitational background field \cite{ADM, V}.

Colladay and Kostelecky \cite{CK} analyzed the question whether Chern-Simons-like 
term is generated by radiative corrections when Lorentz and CPT violating 
term $\bar\psi b_\mu\gamma^\mu\gamma_5\psi$ is added to the conventional 
Lagrangian of quantum electrodynamics in four-dimensional space-time.
They have observed that such term is dependent on regularization scheme.
Coleman and Glashow\cite{CG} argued that such term must unambiguously vanish to 
first order in $b_{\mu}$ for any gauge invariant CPT odd interaction. 
They considered that the axial current $j_{\mu}^5$ should keep gauge invariant 
in the quantum theory at any momentum or at any space-time point. 
Since $\langle j_{\mu}^5 \rangle  = \delta L(x)/\delta b_{\mu}$, this condition is equivalent to 
the requirement that the Lagrangian density corresponding to the quantum effective action
should be gauge invariant. Thus, based on this requirement the Chern-Simons-like term is not
generated since its Lagrangian density is explicitly not gauge invariant. 
Jackiw and Kosteleck\'{y}\cite{jk} 
shown that Chern-Simons-like term is induced. They thought that since $j_{\mu}^5$  only couples 
with a constant 4-vector $b_{\mu}$, it is true to require only that $j_{\mu}^5$ 
with zero-momentum is gauge invariant at quantum level. 
Since $\langle \int d^4x j_{\mu}^5 \rangle = \delta S/\delta b_{\mu}$, this condition is
equivalent to the requirement that the quantum effective action should be gauge invariant. 
This controversy on a possible Chern-Simons-like term generated through
radiative corrections was carefully investigated  by 
many authors \cite{3,4,5,6,7,8,9,10,11,12,13,14}. This phenomena was analyzed in 
quantum electrodynamics as a part of the standard model.
Our purpose in this paper is to derive the Chern-Simons-like gravitational action
induced by Dirac fermions coupled to a background gravitational field. 
 The result show that the Chern-Simons-like term  generated by
radiative fermion loops, under the assumption of the weak field approximation and 
dimensional regularization scheme.

\section{Evaluating the Chern-Simons-like gravitational action}

The action that we are interested is given by
\begin{equation}\label{S1}
S = \int\mathrm{d}^4x (\half i e e^\mu_{\,\,\,\,a} \bar\psi\gamma^a\stackrel{\leftrightarrow}{D}_\mu\psi-e e^\mu_{\,\,\,\,a}\bar\psi b_\mu \gamma^a\gamma_5\psi),
\end{equation}
where we have included the parity-violating term. Here, $e^\mu_{\,\,\,\,a}$ is the tetrad 
(vierbein), $e\equiv\det e^\mu_{\,\,\,\,a}$ and $b_\mu$ is a constant 4-vector 
 . The covariant derivative is given by
\be
D_\mu\psi = \partial_\mu\psi + \half w_{\mu cd}\sigma^{cd}\psi,
\ee
where $w_\mu^{\,\,\,\,cd}$ is the spin connection and $\sigma^{cd} = \frac14[\gamma^c, \gamma^d]$, whereas the covariant derivative on a Dirac-conjugate field $\bar\psi$ is
\be
D_\mu\bar\psi = \partial_\mu\bar\psi - \half w_{\mu cd}\bar\psi\sigma^{cd}.
\ee
Using the expressions above we can rewrite the Eq. (\ref{S1}) as follow
\be
S = \int\mathrm{d}^4x (\half i e e^\mu_{\,\,\,\,a} \bar\psi\gamma^a\stackrel{\leftrightarrow}{\partial}_\mu\psi + {\textstyle{1\over 4}} i e e^\mu_{\,\,\,\,a}  \bar\psi w_{\mu cd}\Gamma^{acd}\psi - e e^\mu_{\,\,\,\,a}\bar\psi b_\mu \gamma^a\gamma_5\psi),
\ee
where $\Gamma^{acd}=\frac16(\gamma^a\gamma^c\gamma^d \pm permutations)$, i.e. the antisymmetrized product of three $\gamma$-matrices.

In the weak field approximation we consider 
$g_{\mu\nu}= \eta_{\mu\nu} + h_{\mu\nu}$ ($g^{\mu\nu}= \eta^{\mu\nu} - h^{\mu\nu}$),
which induces an expansion for the vierbein 
$e_{\mu a}= \eta_{\mu a} + \frac{1}{2}h_{\mu a}$ ($e^\mu_{\,\,\,\,a}= 
\eta^\mu_{\,\,\,\,a} - \frac{1}{2}h^\mu_{\,\,\,\,a}$). Then, the linearized 
Chern-Simons-like action takes the form\cite{JP} 
\be \label{slinear}
S_{linear}= \frac14\int\:d^4x h^{\mu\nu}v^\lambda\epsilon_{\alpha\mu\lambda\rho}\partial^\rho(\partial_\gamma\partial^\gamma h_\nu^\alpha-\partial_\nu\partial_\gamma h^{\gamma\alpha}),
\ee
The main purpose of the present work is to induce this action by radiative correction
of fermionic matter field to obtain the relation between $v_{\lambda}$ and $b_{\mu}$.
In order to perform this
calculation we consider the fermionic model represented by the action 

\be
e^{i\Gamma[h]} = \int \mathcal{D}\bar\psi \mathcal{D}\psi\, e^{iS[h, \bar\psi, \psi]},
\ee
where the linearized effective action is given by
\be\label{secferm}
S[h, \bar\psi, \psi]=\int\mathrm{d}^4x(\half i\bar\psi\Gamma^\mu\stackrel{\leftrightarrow}{\partial}_\mu\psi+\bar\psi h_{\mu\nu}\Gamma^{\mu\nu}\psi-\bar\psi b_\mu\gamma^\mu\gamma_5 \psi),
\ee
with $\Gamma^\mu=\gamma^\mu-\frac12 h^{\mu\nu}\gamma_\nu$ and 
$\Gamma^{\mu\nu}=\frac12 b^\mu\gamma^\nu\gamma_5-\frac{i}{16}(\partial_\rho h_{\alpha\beta})\eta^{\beta\nu}\Gamma^{\rho\mu\alpha}$. 
In this expression, we neglect the terms proportional to $h=\eta^{\mu\nu}h_{\mu\nu}$ 
because they do not contribute to generate the Chern-Simons-like action.

The Feynman rules that we obtain from Eq.(\ref{secferm}) are:\newline 
Fermion propagator
\ben
\raisebox{-0.4cm}{\incps{ferm.eps}{-1.5cm}{-.5cm}{1.5cm}{.5cm}}
 &=& S(p)=\frac{i}{\ps-m}, 
\een
Fermion propagator with $\bs$ insertion
\ben
\raisebox{-0.4cm}{\incps{fermm.eps}{-1.5cm}{-.5cm}{1.5cm}{.5cm}}
&=&  - i \bs \gamma_5.
\een
The three relevant interaction fermion-graviton vertices are
\ben
\raisebox{-.5cm}{\incps{vert.eps}{-1.5cm}{0cm}{1.5cm}{1cm}} &=& -{\textstyle{i\over 4}}\gamma_\mu(2p+q)_\nu,
\een
\ben
\raisebox{-.5cm}{\incps{vertgam.eps}{-1.5cm}{0cm}{1.5cm}{1cm}} 
&=& i\gamma_\mu b_\nu\gamma_5,
\een
and
\ben
\raisebox{-.5cm}{\incps{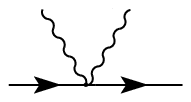}{-1.5cm}{0cm}{1.5cm}{1cm}} &=& -{\textstyle{i\over16}}\eta^{\beta\nu}\Gamma^{\mu\rho\alpha}(q_1-q_2)_\rho.
\een

The one-loop order correction to the effective action 
will be given by the graphs below.
\begin{figure}[h]
\centering
\begin{tabular}{ccccc}
\fgl{(a)}\includegraphics[width=0.15\textwidth]{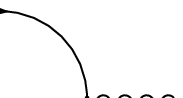} &
\fgl{(b)}\includegraphics[width=0.15\textwidth]{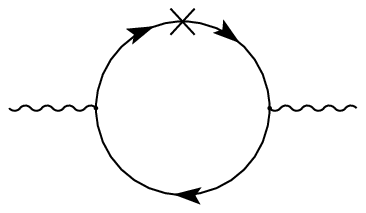} &
\fgl{(c)}\includegraphics[width=0.15\textwidth]{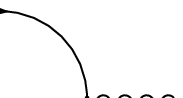} &
\fgl{(d)}\includegraphics[width=0.15\textwidth]{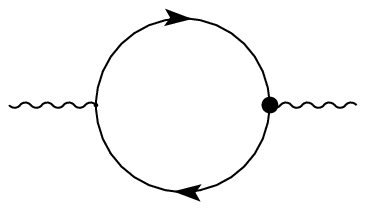} &
(e)\includegraphics[width=0.08\textwidth]{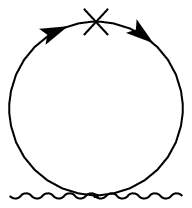}
\end{tabular}
\caption{One-loop contributions}
\label{oneloop}
\end{figure}

The graphs (c), (d) and (e) do not contribute to generate Chern-Simons-like action.
The only relevant graphs are (a) and (b) whose the Feynman integral are given by
\begin{equation}\label{pi1}
\Pi_{a}^{\mu\nu\alpha\beta}(q) = -\frac{i}{16}\,\mathrm{tr} \int\frac{\mathrm{d}^4p}{(2\pi)^4} \left[\gamma^\mu(2p+q)^\nu S(p)\gamma^\alpha(2p+q)^\beta S(p+q)\Slash{b}\gamma_5 S(p+q)\right]
\end{equation}
and
\begin{equation}\label{pi2}
\Pi_{b}^{\mu\nu\alpha\beta}(q) = -\frac{i}{16}\,\mathrm{tr} \int\frac{\mathrm{d}^4p}{(2\pi)^4} \left[\gamma^\mu(2p+q)^\nu S(p)\Slash{b}\gamma_5 S(p)\gamma^\alpha(2p+q)^\beta S(p+q)\right],
\end{equation}.

It is straightforward to see that 
\be
\Pi^{\mu\nu\alpha\beta}(q)=\Pi_{a}^{\mu\nu\alpha\beta}(q)=\Pi_{b}^{\alpha\nu\mu\beta}(-q).
\ee
which appears of substituting the when loop momenta, $p\rightarrow p-qx$, and we using the cyclic 
properties of the trace of a product of $\gamma$-matrices. 
So, from now on we work only with Eq. (\ref{pi1}) which takes the form
\begin{eqnarray}\label{pib1}
\Pi^{\mu\nu\alpha\beta}(q) &=& -\frac{1}{8}\int_0^1\mathrm{d}x\,x\, \int\frac{\mathrm{d}^4p}{(2\pi)^4} \frac{(2p+q(1-2x))^\nu (2p+q(1-2x))^\beta}{[p^2-m^2+x(1-x)q^2]^3} \nonumber\\
&&\times\mathrm{tr}\left[\gamma^\mu(\,\Slash{p}-\Slash{q}x+m)\gamma^\alpha(\,\Slash{p}+\Slash{q}(1-x)+m)\Slash{b}\gamma_5(\,\Slash{p}+\Slash{q}(1-x)+m)\right].
\end{eqnarray}
where we have used Feynman parameter to combine the denominator in  Eq. (\ref{pib1}). First of all, to simplify the numerator of Eq. (\ref{pib1}) 
we take into account that the trace of an odd product of $\gamma$ matrices times
$\gamma_5$ is zero. Thus we have
\ben
\label{pi5}
\Pi^{\mu\nu\alpha\beta}(q) &=& -\frac{1}{8}\int_0^1\mathrm{d}x\,x\, \int\frac{\mathrm{d}^4p}{(2\pi)^4} \frac{(2p+q(1-2x))^\nu (2p+q(1-2x))^\beta}{[p^2-m^2+x(1-x)q^2]^3} \nonumber\\
&&\,\times\mathrm{tr}\left[(\,\Slash{p}+\Slash{q}(1-x))\gamma^\mu(\,\Slash{p}-\Slash{q}x)\gamma^\alpha(\,\Slash{p}+\Slash{q}(1-x))\Slash{b}\gamma_5+m^2(\,\Slash{p}+\Slash{q}(1-x))\gamma^\mu\gamma^\alpha\Slash{b}\gamma_5\right. \nonumber\\
&&+\left.m^2\gamma^\mu(\,\Slash{p}-\Slash{q}x)\gamma^\alpha\Slash{b}\gamma_5+m^2\gamma^\mu\gamma^\alpha(\,\Slash{p}+\Slash{q}(1-x))\Slash{b}\gamma_5\right].
\een
Another algebraic properties of the $\gamma$-matrices 
are used to calculate the numerator of the Eq. (\ref{pi5}).
For instance, we can use that the trace of $\gamma_5$ times an even number of $\gamma$-matrices
can be reduced: 
$Tr[\g^{\mu}\g^{\nu}\g^{\alpha}\g^{\beta}\g_5]=4i\epsilon^{\mu\nu\alpha\beta}$ and 
$Tr[\g^\mu\g^\nu\g_5] = Tr[\g_5]=0$. Also, we dropp all terms that are odd in $p$ to get  
\ben
\label{pi6}
\Pi^{\mu\nu\alpha\beta}(q)&=&-\frac{1}{8}\int_{0}^{1}dx\,x\int\frac{d^4p}{(2\pi)^{4}}
\frac{N^{\mu\nu\alpha\beta}(p^0, p^2, p^4)}{[p^{2}-m^{2}+x(1-x)q^{2}]^{3}}.
\een
When the numerator $N^{\mu\nu\alpha\beta}(p^0, p^2, p^4)$ has the form,
\ben
N^{\mu\nu\alpha\beta}(p^0, p^2, p^4)&=&4p^{\nu}p^{\beta}(T_{0}^{\alpha\mu}+T_{pp}^{\alpha\mu})+2(1-2x)(p^{\nu}q^{\beta}+p^{\beta}q^{\nu})(T_{p}^{\alpha\mu}+T_{ppp}^{\alpha\mu})\nonumber\\
&+&(1-2x)^{2}q^{\nu}q^{\beta}(T_{0}^{\alpha\mu}+T_{pp}^{\alpha\mu})
\een
with
\ben
T_{0}^{\alpha\mu}&=&-4ib_{\lambda}\epsilon^{\alpha\mu\lambda\theta}q_{\theta}[x(1-x)^2q^2+(2-x)m^{2}]\\
T_{p}^{\alpha\mu}&=&-4ib_{\lambda}\epsilon^{\alpha\mu\lambda\rho}[m^{2}+(1-x^2)q^{2}]p_{\rho} \nonumber \\
&&-8i(1-x)b_{\lambda}[\epsilon^{\mu\lambda\rho\theta}q^{\alpha}-
\epsilon^{\alpha\lambda\rho\theta}q^{\mu}-(1-x)\epsilon^{\alpha\mu\lambda\theta}q^{\rho}]q_{\theta}p_{\rho}\\
T_{pp}^{\alpha\mu}&=&-4ib_{\lambda}[2(\epsilon^{\mu\lambda\rho\theta}p_{\rho}p^{\alpha}-\epsilon^{\alpha\lambda\rho\theta}p_{\rho}p^{\mu}+x \epsilon^{\alpha\mu\lambda\rho}p_\rho p^\theta)-(2-x)\epsilon^{\alpha\mu\lambda\theta}p^{2}]q_{\theta}\\
T_{ppp}^{\alpha\mu}	&=&4ib_{\lambda}\epsilon^{\alpha\mu\lambda\rho}p^{2}p_{\rho}
\een

The integral (\ref{pi6}) is badly divergent. By power counting we note that the
kind of divergences are quadratic and logarithmic. Pehaps the mostconvenient method for 
regulating divergent integrals without impairing gauge invariance is the 
dimensional regularization scheme developed by 't Hooft and Veltman \cite{thooft} in 1972.   
Thus, we change dimensions from $4$ to $D$ and
we change $d^4p/(2\pi)^4$  to $(\mu^2)^{(2-D/2)}[d^{D}p/(2\pi)^{D}]$, where $\mu^2$ is 
an arbitrary parameter that identifies the mass scale. Thus Eq. (\ref{pi5}) takes the form
\begin{equation}
\label{pib3}
\Pi^{\mu\nu\alpha\beta}(q) = b_\lambda\epsilon^{\alpha\mu\lambda\rho}q_\rho\left[A\,q^2 \eta^{\beta\nu} + 
B\,q^\beta q^\nu \right],
\end{equation}
where $A$ and $B$ are given by
\begin{equation}
\label{pill}
A = \frac{-1}{32\pi^2}\int_0^1\mathrm{d}x\,\left[(3-2x)x^2(1-x) \Gamma(\epsilon/2)-3x^2\frac{M^2}{q^2}\Gamma(-1+\epsilon/2)\right]\left(\frac{4\pi\mu^2}{-M^2}\right)^{\epsilon/2}
\end{equation}
and
\ben
\label{pilq}
B &=& \frac{-2}{32\pi^2}\int_0^1\mathrm{d}x\,(1-2x)x^2(1-x)\Gamma(\epsilon/2)\left(\frac{4\pi\mu^2}{-M^2}\right)^{\epsilon/2}\\
&+& \frac{1}{64\pi^2}\int_0^1\mathrm{d}x\,(1-2x)^2\left[x(2-3x)+(3-2x)x^2(1-x)\frac{q^2}{M^2}\frac{\epsilon}{2}\right]\Gamma(\epsilon/2)\left(\frac{4\pi\mu^2}{-M^2}\right)^{\epsilon/2},
\een
where $\epsilon=4-D$ and $M^2=m^2-x(1-x)q^2$. The next step concern expanding  the 
gamma function that appear in $A$ and $B$ around 
$\epsilon \to 0$, thus we have
\begin{eqnarray}\label{pille}
A &=& \frac{1}{32\pi^2}\int_0^1\mathrm{d}x\,\left[3x^3(1-x)+(5x-3)x^2(1-x)\left(\frac{2}{\epsilon}+\ln\left(\frac{4\pi\mu^2}{-M^2}\right)-\gamma\right)\right. \nonumber\\&-&\left.3x^2\left(\frac{2}{\epsilon}+\ln\left(\frac{4\pi\mu^2}{-M^2}\right)-\gamma+1\right)\frac{m^2}{q^2}\right]
\end{eqnarray}
and
\begin{eqnarray}
B&=& \frac{1}{64\pi^2}\int_0^1\mathrm{d}x\,\left\{(1-2x)^2(3-2x)\frac{x^2(1-x)q^2}{M^2}\right. \nonumber\\
&+&\left.\left[(2-3x)(1-2x)-4x(1-x)\right]x(1-2x)\left(\frac{2}{\epsilon}+\ln\left(\frac{4\pi\mu^2}{-M^2}\right)-\gamma\right)\right\}.
\end{eqnarray}
As one can see $\int_0^1dx[(5x -3)x^2(1-x)](\frac{2}{\epsilon} - \gamma)=0$ in $A$ 
and $\int_0^1dx[(2-3x)(1-2x) - 4x(1-x)]x(1-2x)(\frac2{\epsilon}-\gamma)=0$ in 
$B$, then $A$ and $B$ take the form
\begin{eqnarray}
A &=& \frac{1}{32\pi^2}\int_0^1\mathrm
{d}x\,\left[3x^3(1-x)+\frac{(1-2x)x^3(1-x)^2q^2}{m^2-x(1-x)q^2}
\right. \nonumber \\ &-& \left.3x^2\left(\frac{2}{\epsilon}+\ln\left(\frac{4\pi\mu^2}{-M^2}\right)-\gamma+1\right)\frac{m^2}{q^2}\right]
\end{eqnarray}
and
\begin{eqnarray}
B &=& \frac{1}{32\pi^2}\int_0^1\mathrm{d}x\,\frac{(1-2x)^2x^2(1-x)q^2}{m^2-x(1-x)q^2}
\end{eqnarray}
Observe that we have performed an integration by parts on $x$ for log term in $A$ and $B$.
Note that in $A$ the divergent part is present which will disappear when we consider
the limit $m^2\rightarrow 0$. Now performing the $x$-integration, we have
\begin{equation}\label{pi_general}
A\stackrel{m^2\rightarrow 0}{=} - B\stackrel{m^2\rightarrow 0}{=} \frac{1}{192\pi^2}.
\end{equation}
We use these results into Eq. (\ref{pib3}), to obtain the Chern-Simons-like term 
\begin{equation}
\Pi^{\mu\nu\alpha\beta}(q) = \frac{1}{192\pi^2}b_\lambda\epsilon^{\alpha\mu\lambda\rho}q_\rho\left[\,q^2 \eta^{\beta\nu} - 
\,q^\beta q^\nu \right],
\end{equation}
Finally, the Chern-Simons-like gravitational action induced by
radiatively corrections is given by
\be
\label{sf}
\Gamma_{\mathrm{cs}}[h] = \frac{1}{192\pi^2} \int\mathrm{d}^4x b^\lambda h^{\mu\nu}\epsilon_{\alpha\mu\lambda\rho}\partial^\rho\left[\partial_\gamma\partial^\gamma h^\alpha_\nu-\partial_\nu\partial_\gamma h^{\gamma\alpha}\right],
\ee
Comparing to Eq.(\ref{slinear}) we obtain the relation between the parameters 
$v_{\lambda}$ and $b_{\mu}$ which is written as 
\be
v_\lambda=\frac{1}{48\pi^2}b_\lambda.
\ee
\section{conclusions}
We summarize our work recalling that we have calculated the radiative corrections induced
by Dirac fermions coupled to gravitational background field, including the
nonstandard contribution $\bar\psi \bs \gamma_5\psi$ that violate parity symmetries.
In this calculation we have used the weak field approximation and dimensional regularization 
scheme. The coefficient 
of the Chern-Simons-like gravitational action obtained in  Eq. (\ref{sf}) is in agreement
with the result obtained by Avarez-Gaum\'e and  Witten within the context of gravitational 
anomalies\cite{aw}. 

\section*{Acknowledgments}

We would like to thank D. Bazeia for useful discussions, CAPES, CNPq and PROCAD for partial support.

\end{document}